\documentstyle[prl,aps,multicol]{revtex}

\begin{document}

\title{ Comment on ``Protective measurements of the wave function of a
  single squeezed harmonic-oscillator state''}
\author{Y. Aharonov$^{a,b}$,  and L. Vaidman$^a$}

\address{$^a$ School of Physics and Astronomy,
Raymond and Beverly Sackler Faculty of Exact Sciences, \\
Tel-Aviv University, Tel-Aviv 69978, Israel.\\
$^b$ Physics Department, University of South Carolina,\\
Columbia, South Carolina 29208, USA}

\date{}

\maketitle

\begin{abstract}
  Alter and Yamamoto [Phys. Rev. A {\bf 53}, R2911 (1996)] claimed to
  consider ``protective measurements'' [Phys. Lett. A {\bf 178}, 38
  (1993)] which we have recently introduced. We show that the
  measurements discussed by Alter and Yamamoto {\em are not} the
  protective measurements we proposed. Therefore, their results are
  irrelevant to the nature of protective measurements.
\end{abstract}

\begin{multicols}{2}

In a recent Rapid Communication  Alter and Yamamoto \cite {orly}
considered  sequences of certain measurements of a squeezed
harmonic-oscillator state. They claimed that these are the {\em
  protective measurements} \cite{AV,AAV} which we have recently
proposed for observing the wave function a single quantum system.
While we do not want to make any statement about the significance of the
measurements discussed by Alter and Yamamoto, we claim that these
measurements {\em are not} the protective measurements we proposed.
Therefore, all conclusions of the authors about protective
measurements are unfounded and, in particular, their central result
that ``The protective measurement requires a full {\em a priori}
knowledge of the measured wave function'' is incorrect.

There are two basic ingredients in the protective measurements we have
proposed. The first is that the quantum state of the system is
protected, i.e., it is a non-degenerate energy eigenstate with a
finite gap to a neighbor level. The second is that the interaction is
not infinitely fast and strong as in ideal measurements, but slow and
weak enough  that the adiabatic approximation is applicable (the
probability that the system leaves the energy eigenstate is
negligible) and  the state does not changed significantly during
the measurement, and therefore the measurement shows the property of
the observed quantum wave function. 

 Our protective measurements have
also a property that there is almost no entanglement between the
system and the measuring device at the end of the measurement
interaction. It seems that Alter and Yamamoto took the latter together
with the weakness of the measurement coupling as the definition of
a protective measurement, completely ignoring the first basic ingredient
of the protective measurements, the protection.  The squeezed
harmonic-oscillator state on which 
measurements  are described in the Rapid Communication, {\em is
  not protected}! It is not an energy eigenstate and it is not
protected by frequent observations -- an alternative protection
procedure based on the quantum Zeno effect. (The process of ``driving the
signal back to its initial excitation'' described in the Rapid
Communication does not entail the Zeno effect.)

The property that ``the signal and the  probe are left disentangled
after their interaction'' is also the property of  ``ideal von
Neumann 
measurements'' \cite{vN} which are frequently called ``non-demolition
measurements'' \cite{NDM}. If the initial state of the quantum system is
an eigenstate of some observable,
then an ideal (non-demolition) measurement of this observable  does not
change the quantum state. Weakening the von Neumann coupling does not
change this property. Alter and Yamamoto considered such measurements
with weak and
strong coupling, naming the former ``protective
measurements''. Indeed, their procedure, together with  ``driving the signal back'' falls into
this category. The coupling leads to a known (for a given initial
state) change which is then corrected. Note also the difference
between ``protective measurements'' and ``ideal measurements''
regarding the disentanglement property. If the initial quantum state
is  a protected state, then no adiabatic weak measurement of {\em any}
variable  leads to entanglement between the system and the
measuring device. In contrast, ideal measurements of only the
observables 
for which the initial state of the system is an eigenstate  do not lead to entanglement. 

The quantum state of a harmonic oscillator which can be observed using
protective measurement is one of the energy eigenstates. We do not
need to know the full information about the state: the only
information needed is that it is an energy eigenstate and that the gap
to any other eigenstate is larger than some value. This information is
necessary for fixing the parameters needed for adiabaticity of the
protective measurement.

 If the potential, i.e.,
the strength and the location of the harmonic oscillator, are known,
then the ideal measurement can tell us what is the  energy  of the system and we can calculate the expectation
value of any observable -- the outcome of corresponding protective
measurement. (Moreover, we do not need any information about the energy
gap for the ideal measurement of energy.) But, if we do not know the potential,
the protective measurement can give us more information than any  ideal
measurement can. The bound on the energy gap is the only {\em a priori}
information which is needed for finding the complete wave function of
a nondegenerate energy eigenstate of an unknown potential.  

We  note that our work on protective measurement has been
misinterpreted, although in another way, before \cite{U,R,GH}. We
hope that this comment and our reply \cite {AAV-f}  clarify the
concept of protective measurements. We want also to point out a recent
 generalization of this concept to  protective measurements of   pre- and post-selected systems
\cite{AV-P}  and to  protective measurements of  metastable systems \cite{AMPTV}.

This research was supported in part by  grant 614/95 of the 
Basic
Research Foundation (administered by the Israel Academy of Sciences and
Humanities) and   by grant PHY-9307708 of the National Science
Foundation.

\end{multicols}

\end{document}